\documentstyle[prd,aps]{revtex}

\begin{document}
\preprint{SUSSEX-AST 96/4-1, astro-ph/9604001}
\draft

%
% Remove this and closure after abstract, plus preprint number,
% in electronic submission
%
\input epsf
\twocolumn[\hsize\textwidth\columnwidth\hsize\csname 
@twocolumnfalse\endcsname

\title{Conditions for successful extended inflation}
\author{Anne M. Green and Andrew R. Liddle}
\address{Astronomy Centre, University of Sussex, Falmer, Brighton BN1 
9QH,~~~U.~K.}
\date{\today}
\maketitle
\begin{abstract}
We investigate, in a model-independent way, the conditions required to 
obtain a satisfactory model of extended inflation in which inflation is 
brought to an end by a first-order phase transition. The constraints are 
that the correct present strength of the gravitational coupling is obtained, 
that the present theory of gravity is satisfactorily close to general 
relativity, that the perturbation spectra from inflation are compatible with 
large scale structure observations and that the bubble spectrum produced at 
the phase transition doesn't conflict with the observed level of microwave 
background anisotropies. We demonstrate that these constraints can be 
summarized in terms of the behaviour in the conformally related Einstein 
frame, and can be compactly illustrated graphically. We confirm the 
failure of existing models including the original extended inflation model, 
and construct models, albeit rather contrived ones, which satisfy all 
existing constraints.
\end{abstract}
\pacs{PACS numbers: 98.80.Cq \hspace*{2cm} Sussex preprint SUSSEX-AST 
96/4-1, 
astro-ph/9604001}

\vskip2pc]

%%%%%%%%%%%%%%%%%%%%%%%%%%%%%%%%%%%%%%%%%%%%%%%%%%%%%%%%%%%%%%%%%%%%%%%%
\section{Introduction}

The extended inflation scenario \cite{LS,K91} offers the prospect of 
resuscitating the original idea of Guth \cite{G81} that inflation 
\cite{LINDE,KT,LLrep} could be driven by a metastable vacuum energy and end 
by the tunnelling of the associated scalar field to the true minimum of its 
potential. The strategy is to implement inflation in an extended theory of 
gravity, such as a scalar--tensor theory, in which the expansion rate 
induced by a vacuum energy is slower than exponential. Under such 
circumstances, one is guaranteed that the phase transition will be able to 
successfully complete, which proves not to be the case in Einstein gravity 
if sufficient inflation is demanded to solve the usual cosmological 
problems.

Moving to an extended gravity theory is an interesting way of generalizing 
existing inflation models, because such scenarios are much more highly 
constrained \cite{W93} than alternative generalizations where extra scalar 
fields are added by hand. In particular, one knows that the theory must 
mimic general relativity to a high degree at the present epoch, and there 
are further strong constraints on the variation of the strength of the 
gravitational interaction going all the way back to the time of 
nucleosynthesis. These additional constraints naturally have the effect of 
making it more difficult to obtain a viable model. 

When one chooses to end inflation via a first-order phase transition, where 
bubbles of true vacuum nucleate, expand and coalesce, this introduces 
further constraints, because it is possible for the earliest true-vacuum 
bubbles which nucleate to be caught up in the subsequent inflationary 
expansion and stretched to astrophysically large sizes \cite{W89,LSB}. These 
can contribute both density perturbations and microwave background 
anisotropies over and above those caused by quantum fluctuations 
\cite{PERTEI} as in all inflationary models \cite{PERT}. Near the general 
relativity limit, the distribution of bubbles is scale-invariant (in the 
sense of equal volume residing in bubbles within a given logarithmic size 
interval) which is far from acceptable \cite{W89,LSB}.

The original extended inflation model \cite{LS} was implemented in the 
Jordan--Brans--Dicke (JBD) theory of gravity \cite{BD,W93}, where the 
gravitational `constant' is replaced by a field $\Phi$ whose variation is 
controlled by a coupling parameter $\omega$. General relativity is obtained 
in the limit of large $\omega$. In that model, it was quickly shown that the 
competing needs of staying close to the general relativity limit to match 
present observations ($\omega>500$ \cite{Retal,W93}), and of obtaining a 
satisfactory bubble distribution ($\omega \lesssim 25$ \cite{W89,LSB}), are 
mutually exclusive. This became known as the big-bubble problem, and various 
strategies have been brought into play in an attempt to evade it. The 
simplest is to introduce a mechanism which invalidates the present-day 
bound on $\omega$; this can for example be achieved by introducing a 
potential for the Brans--Dicke field $\Phi$ which is negligible during 
inflation and which prevents its variation at the present epoch. 
Alternatively, one can move to a general scalar--tensor theory, in which 
$\omega$ is allowed to depend on $\Phi$, which allows one to exercise 
control over how closely the general relativity limit is attained at 
different epochs.

Both these strategies have more recently suffered further constraints, 
under the assumption that the quantum fluctuations during inflation provide 
the density perturbations which are responsible for large-scale structure 
and microwave background anisotropies. The results of COBE in combination 
with large-scale structure studies quickly led to the conclusion that the 
spectrum of density perturbations must not be too far from scale-invariant, 
if the observed structures are to be reproduced. However, in extended 
inflation models one expects that if one makes the necessary moves to {\em 
break} the scale-invariance of the bubble distribution, then one will also 
destroy the scale-invariance of the density perturbation spectrum. This 
implies two opposing constraints, but now both to be applied during 
inflation. This extra consideration proved sufficiently stringent to exclude 
all models in the existing literature in which inflation ends by 
nucleation \cite{LL,LLEI}\footnote{There are however models such as 
hyperextended inflation \cite{SA}, in which inflation is brought to an end 
through dynamical evolution, with bubbles then nucleating in the 
post-inflationary phase.}.

In the literature, a substantial number of models falling into the extended 
inflation class have been devised \cite{MODELS,SA,BM}, and examined on a 
more or less case by case basis \cite{Yun,LW2}. In this paper, we shall 
place the constraints in a more general framework, allowing one to see 
easily the problems of existing models. As a by-product, this will enable us 
to construct working models satisfying all present constraints, though as we 
shall see the constraints combine in such a way as to make such models 
appear extremely contrived. The prognosis for the extended inflation 
scenario therefore continues to look poor.

\section{Extended inflation models}

We work within a class of theories featuring a general scalar--tensor theory 
with Brans--Dicke field $\Phi$, plus a separate scalar field $\sigma$ 
trapped 
in a metastable state with energy density $V_0$. Models of this sort were 
discussed in Refs.~\cite{BM,LW2}. The action is
\begin{eqnarray}
S & = & \int \! {\rm d}^4 x \sqrt{-g} \left\{ \frac{m_{{ \rm Pl }}^2}{16\pi}
	\left[ \Phi R - \frac{\omega(\Phi)}{\Phi} \partial_{\mu} \Phi
	\partial^{\mu} \Phi + 2 \Phi \lambda(\Phi) \right] \right. 
	\nonumber \\
 & & \quad \quad - \left. \frac{1}{2} \partial_{\mu} \sigma \partial^{\mu} 
 	\sigma + V(\sigma) \right\} \,.
\end{eqnarray}
We have normalized the Brans--Dicke field by pulling out a pre-factor; with 
these conventions $\Phi$ is dimensionless and its present value is 1. The 
function $\lambda(\Phi)$ corresponds to a potential for the Brans--Dicke 
field, which we shall assume to be negligible during inflation. We need 
assume nothing about the potential for $\sigma$ beyond that the metastable 
state has potential $V_0$ and that the field can tunnel from this to the 
true vacuum.

A vital tool for studying this theory is the conformal transformation 
\cite{M89}, which simplifies the gravitational sector to general relativity. 
On performing the conformal transformation $g_{\mu \nu} = \Omega^{-2} 
\tilde{g}_{\mu \nu}$ with $\Omega = \sqrt{\Phi}$, and defining a new scalar 
field by
\begin{equation}
\label{phiPhi}
\phi \equiv \frac{m_{{\rm Pl}}}{\sqrt{8\pi }} \int 
	\frac{{\rm d} \Phi}{\Phi} \sqrt{ \omega(\Phi) +\frac{3}{2}} \,,
\end{equation}
one obtains the equivalent action 
\begin{eqnarray}
S & = & \int {\rm d}^4 x \sqrt{-\tilde{g}} \left[ 
	\frac{m_{{\rm Pl}}^2}{16\pi } \tilde{R}- \frac{1}{2}
	\partial_{\mu}\phi\partial^{\mu}\phi \right. \nonumber \\
 & & \left. \quad \quad  + 2 U(\phi) - \frac{1}{2\Omega^{2}} 
	\partial_{\mu} \sigma \partial^{\mu} \sigma +
	\frac{V(\sigma)}{\Omega^{4}} \right] \,,
\end{eqnarray} 
where $U(\phi) \equiv \lambda(\Phi) m_{{\rm Pl}}^2/16 \pi \Phi$. Here 
`tilde' indicates quantities in the transformed frame, which we refer to as 
the Einstein frame (the original one being the Jordan frame). The 
field $\phi$ has been defined so as to have a canonical kinetic 
term. Assuming 
$\lambda(\Phi)$ vanishes and that the $\sigma$ field is in its trapped 
phase, this simplifies to
\begin{equation}
S = \int {\rm d}^4 x \sqrt{-\tilde{g}} \left[ 
	\frac{m_{{\rm Pl}}^2}{16\pi } \tilde{R}- \frac{1}{2}
	\partial_{\mu}\phi\partial^{\mu}\phi + V_0 \Omega^{-4} \right] \,.
\end{equation}
This is the action for a standard single field chaotic inflation model 
\cite{LINDE} with potential
\begin{equation}
\label{VPhi}
V(\phi) = V_0 \Omega^{-4} = V_0 \Phi^{-2} \,,
\end{equation} 
where $\Phi(\phi)$ is given by Eq.~(\ref{phiPhi}). A change in the 
coupling function $\omega(\Phi)$ in the Jordan frame therefore leads to a 
different potential $V(\phi)$ in the Einstein frame, by changing the 
relation between $\phi$ and $\Phi$. Equally, if we are given a potential 
$V(\phi)$ in the Einstein frame, there exists a corresponding Jordan frame 
theory with a trapped field $\sigma$ and a coupling function $\omega(\Phi)$. 
The only difference is that in the standard chaotic inflation scenario 
inflation ends when the potential becomes too steep to maintain inflation, 
while in our scenario inflation may also end when the $\sigma$ field 
tunnels, which can happen at any location on the potential $V(\phi)$.

Our strategy is to use the Einstein frame scalar field $\phi$ as a time 
variable, even when referring to quantities in the Jordan frame. The bulk of 
the analysis shall be carried out in the Einstein frame, using the usual 
slow-roll approximation. In the Einstein frame, we define the slow-roll 
parameters \cite{LL}
\begin{eqnarray}
\label{eps}
\epsilon(\phi) & = & \frac{m_{{\rm Pl}}^2}{16 \pi} \left( 
	\frac{V'}{V} \right)^2 \,; \\
\eta(\phi) & = & \frac{m_{{\rm Pl}}^2}{8 \pi} \frac{V''}{V} \,,
\end{eqnarray}
where primes indicate derivatives with respect to $\phi$. Inflation 
occurs in the Einstein frame provided these are much less than one. The 
number of $e$-foldings of inflation in the Einstein frame, between two 
values of $\phi$, is given by the usual formula
\begin{equation}
\label{N}
N(\phi_1,\phi_2) \simeq - \frac{8 \pi}{m_{{\rm Pl}}^2}
	\int_{\phi_1}^{\phi_2} \frac{V}{V'} \, {\rm d}\phi \,.
\end{equation}

From Eqs.~(\ref{phiPhi}) and (\ref{VPhi}), one can derive \cite{GBW} a 
remarkably simple relation between $V(\phi)$ and $\omega(\Phi)$
\begin{equation}
\label{omeps}
\omega(\Phi) + \frac{3}{2} = \frac{2}{\epsilon(\phi)} \,.
\end{equation}
In the particular case of the JBD theory where $\omega$ is constant, then 
$\epsilon$ is constant too and we have the well known result of power-law 
inflation in the Einstein frame.

\section{Observational constraints}

Here we go through the constraints on the scenario.

\subsection{Present strength of gravity}

To reproduce the present value of the gravitational coupling, we require 
$\Phi = 1$ at the present. If $\Phi$ is a free field, then the present 
strength of gravity is determined dynamically, by whatever $\Phi$ happens to 
have evolved to. During conventional evolution such as matter domination or 
radiation domination, the variation of $\Phi$ is extremely slow 
\cite{N69,W89}, unless $\omega$ is very small. Consequently, it is an 
excellent approximation to assume that the present value is the same as that 
at the end of inflation. In that case $V_0$ can be identified with the 
Einstein frame potential energy at the end of inflation.

If instead there is a potential for $\Phi$, its value at the end of 
inflation becomes unimportant and instead we require that the minimum  of 
the potential be at $\Phi = 1$; the present gravitational coupling is 
therefore determined by the parameters of the theory. In that case, the 
value of $\Phi$ at the end of inflation can effectively be incorporated into 
the amplitude of $V(\phi)$, by writing
\begin{equation}
V(\phi) = \frac{V_0}{\Phi_{{\rm end}}^2} \left(
	\frac{\Phi_{{\rm end}}}{\Phi} \right)^2 \,.
\end{equation}
Then the potential energy at the end of inflation in the Einstein frame is 
$V_0/\Phi_{{\rm end}}^2$.

In the following expressions, we shall assume a free $\Phi$ field so that 
$V(\phi_{{\rm end}}) = V_0$. The introduction of a potential for $\Phi$ can 
be simply accounted for via the scaling above.

\subsection{Recovering general relativity}

If the $\Phi$ field is a free field, then present-day limits from solar 
system observations demand that $\omega >500$ \cite{Retal,W93}, with a 
comparable constraint from nucleosynthesis \cite{NUCL}. (There is 
also a limit on ${\rm d}\omega/{\rm d}\Phi$, but typically this is much 
weaker.) This constraint is extremely difficult to satisfy within the 
extended inflation context --- we see from Eq.~(\ref{omeps}) that it 
requires $\epsilon(\phi)$ at the end of inflation to be less than $1/250$. 
This tough constraint can be evaded by the introduction of a potential for 
the $\Phi$ field, which prevents its variation at the low energy scales of 
our present universe \cite{LSB} and permits any present value of $\omega$.

\subsection{The bubble spectrum}

Bubble nucleation is typically discussed in the Jordan frame. The nucleation 
rate per unit volume per unit time $\Gamma$ is a constant determined by the 
shape of the potential barrier between the false and true vacuum states 
(see e.~g.~Ref.~\cite{KT}), but the quantity of interest is not this, but 
rather the nucleation rate per Hubble volume per Hubble time
\begin{equation}
E = \frac{\Gamma}{H^4} \,,
\end{equation}
where $H$ is the Jordan frame Hubble parameter. Once $E$ exceeds some 
critical value of order unity, the phase transition is able to complete 
\cite{GW}. We shall assume that the critical value is unity, in which case 
$\Gamma = H_{{\rm end}}^4$.

The bubbles which are potentially constraining are those which are stretched 
to large sizes by the subsequent inflationary expansion, so they would be 
nucleated some time before the end of inflation. For typical parameters, a 
bubble of present size $20 h^{-1}$ Mpc would have nucleated about $55$ 
$e$-foldings from the end\footnote{It doesn't really matter in which frame 
the $e$-foldings are computed; see the Appendix.}. The crucial quantity is 
therefore the nucleation rate at that time, which we shall denote by 
$E_{55}$. Original calculations of the bubble spectrum \cite{W89,LSB,Wu} 
were combined with fairly {\it ad hoc} observational criteria to obtain the 
constraint $\omega \lesssim 25$ on the original extended inflation model. 
These calculations were followed up by more specific ones \cite{LW} which 
included detailed calculations of the effect of bubbles on the microwave 
background; it was estimated that any bubble larger than $20 h^{-1}$ Mpc 
would be seen in the microwave background, and integration over the bubble 
spectrum leads to a slightly stronger version of the constraint.

However, it is possible to adopt a much more straightforward approach which 
captures the essence of the constraint, which is simply to limit the value 
of $E$ at the time the dangerous bubbles are forming. Because $E$ is growing 
with time, the most constraining bubbles are always the smallest ones that 
you can see, and so it is reasonable to take the constraint as being on the 
instantaneous value of $E$ when those bubbles formed, i.e. $E_{55}$. This 
approach was adopted in Ref.~\cite{CLLSW}, where it was shown present 
observations demand
\begin{equation}
E_{{\rm 55}} = \left( \frac{H_{{\rm end}}}{H_{55}} \right)^4 < 10^{-5}\,.
\end{equation}
This constraint is extremely conservative, since many very generous 
assumptions were made in computing the effect of bubbles on the microwave 
background \cite{LW}.

We need to compute the Jordan frame Hubble parameter in terms of the 
Einstein frame quantities. Continuing to use `tilde' to indicate Einstein 
frame quantities, the transformation rule
\begin{equation}
\label{neq1}
{\rm d}t = \Omega^{-1} {\rm d} \tilde{t} \quad ; \quad a(t) = \Omega^{-1}
	\tilde{a}(\tilde{t}) \,,
\end{equation}
implies, continuing to measure time with the Einstein frame scalar field, 
that
\begin{equation}
\label{neq2}
H( \phi ) = \Omega \tilde{H} - \frac{{\rm d}\Omega}{{\rm d}t} \,.
\end{equation}
Some algebra brings us to the result
\begin{equation}
\label{ophi}
 H(\phi) = \left( \frac{8\pi}{3 m_{{\rm Pl}}^2} \right)^{1/2} V_0^{1/4}
	V^{1/4}(\phi) \left[ 1 - \frac{\epsilon(\phi)}{2} \right] \,.
\end{equation}
Notice that the Jordan frame Hubble parameter only goes as the fourth root 
of the Einstein frame potential, whereas the Einstein frame Hubble parameter 
goes as the square root. From this expression we obtain
\begin{equation}
\frac{H(\phi)}{H_{{\rm end}}} = \left( \frac{V(\phi)}{V_0} \right)^{1/4}
	\left[ \frac{1-\epsilon/2}{1-\epsilon_{{\rm end}}/2} \right] \,.
\end{equation}
Unless the $\epsilon$ terms are anomalously large, the square bracketed term 
can be taken as unity, and the bubble constraint translated into the 
Einstein frame becomes simply
\begin{equation}
\label{bubcons}
\frac{V(\phi_{55})}{V_0} \ge 10^5 \,.
\end{equation}

\subsection{Density perturbations}

Because we know that general relativity is a good description of the present 
universe, we know that the Jordan and Einstein frames must coincide to high 
accuracy at the present. Therefore calculations can be carried out in 
whichever frame is easiest. We use the Einstein frame.

Inflation models produce spectra of both density perturbations and 
gravitational waves. It is natural to assume that these perturbations are 
those responsible for structure in the universe, and under that assumption 
their form is strongly constrained by a variety of observations. The general 
scalar--tensor case, where both the inflaton and Brans--Dicke fields exhibit 
fluctuations, leads to a complicated phenomenology \cite{SY,GBW}. Things are 
considerably simpler here, where the $\sigma$ field is trapped, and the 
situation for Jordan--Brans--Dicke theory was established in 
Ref.~\cite{PERTEI}. For the general scalar--tensor scenario we discuss, the 
relevant results can be found as a limiting case of the general analysis of 
Ref.~\cite{GBW}.

The crucial parameters are the amplitude of the density perturbations, 
which, following \cite{LL,LLrep}, we shall denote $\delta_{{\rm H}}$, the 
spectral index $n$ of the density perturbations and the contribution $R$ of 
gravitational waves to large angle microwave background anisotropies. In the 
Einstein frame, these are given by the standard formulae \cite{LL,LLrep}
\begin{eqnarray}
\delta_{{\rm H}} & = & \frac{32}{75} \, \frac{V_*}{m_{{\rm Pl}}^4} \,
	\frac{1}{\epsilon_*} \,; \\
n & = & 1 - 6 \epsilon_* + 2 \eta_* \,; \label{n} \\
R & \simeq & 12.4 \epsilon_* \,,
\end{eqnarray}
where subscript `$*$' indicates that the quantities are to be evaluated when 
the relevant scales crossed outside the Hubble radius during inflation. The 
largest scales (such as the microwave background quadrupole) typically 
correspond to about 60 $e$-foldings from the end of inflation, while the 
shortest scales (corresponding to galaxy formation) are at about 50 
$e$-foldings. So we can take `$*$' to indicate 55 $e$-foldings, the same as 
the bubbles, since the interesting $20 h^{-1}$ Mpc scale for the bubbles is 
roughly in the middle of the large scale structure range.

The overall amplitude of perturbations can be fixed to match observations 
through scaling the potential by a suitable factor, thus determining $V_0$ 
(or $V_0/ \Phi_{{\rm end}}^2$ if there is a potential for $\Phi$). The other 
two parameters can be constrained through compilation of large scale 
structure observations spanning as wide a range of scales as possible. This 
has been recently done for inflationary spectra by Liddle et al. 
\cite{LLSSV}, who considered the case of critical density, allowing 
arbitrary mixtures of cold dark matter and hot dark matter. The constraints 
can be summarized in terms of $\epsilon_{55}$ and $\eta_{55}$, and can be 
taken to be
\begin{equation}
4 \epsilon_{55} - \eta_{55} < 0.20 \quad ; \quad \eta_{55} - 
	\epsilon_{55} < 0.10 \,.
\end{equation}
These are quite conservative, allowing the Hubble constant and the amount of 
hot dark matter to be freely chosen to allow the best possible fit. Specific 
choices for them would strengthen the constraints. The constraints map out a 
small area in the $(\epsilon_{55},\eta_{55})$ plane; in particular the 
maximum value of $\epsilon_{55}$ permitted by them is 0.10 (at $\eta_{55} = 
0.20$); for any higher value the amount of gravitational waves in the COBE 
signal is so high as to render the density perturbations too weak to give 
the observed structures.

\section{Graphical interpretation of the constraints}

We have found that the best way to illustrate the competing nature of these 
constraints is graphically, by plotting $\ln V$ against $N$, where $N$ is 
the number of $e$-foldings before the end of inflation. The reason is 
that one can rewrite $\epsilon$, using Eq.~(\ref{N}), as
\begin{equation}
\label{epsV}
\epsilon = \frac{1}{2} \frac{{\rm d} \ln V}{{\rm d}N} \,,
\end{equation}
so $\epsilon$ simply gives the gradient of $\ln V(N)$. Furthermore, the 
curves corresponding to the original extended inflation model are straight 
lines in such a graph.

The bubble constraint immediately lets us constrain the mean $\epsilon$ over 
the last 55 $e$-foldings, since
\begin{equation}
\bar{\epsilon} \equiv \frac{1}{2} \frac{\Delta \ln V}{\Delta N} > 0.10 \,,
\end{equation}
according to Eq.~(\ref{bubcons}). This immediately shows the problem with 
the original extended inflation model; since $\epsilon$ is constant 0.10 is 
its minimum value, in conflict with the large scale structure requirement. 
Although as we have quoted it this seems very marginal, one should bear in 
mind that both sets of constraints are extremely conservative, especially 
the bubble one. A more accurate computation would create a significant gap 
between the two requirements.

\begin{figure}
\centering 
\vspace*{0.2cm}
\leavevmode\epsfysize=6.9cm \epsfbox{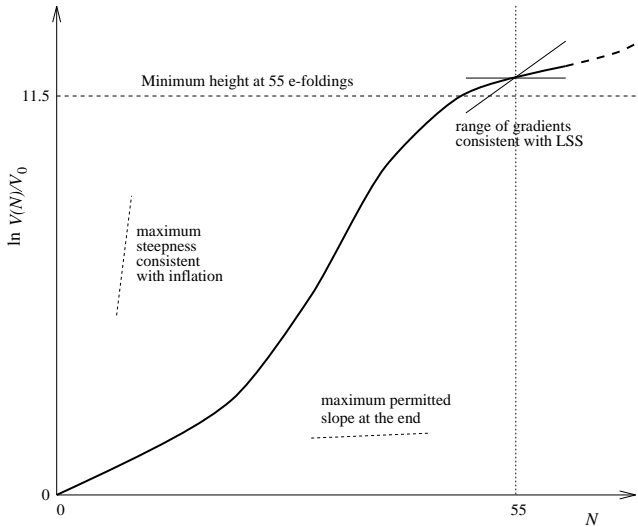}\\ 
\vspace*{0.2cm}
\caption[constr]{\label{constr} A schematic diagram of the constraints, with 
gradients plotted roughly to scale. The short solid lines indicate the 
allowed gradient of the $\ln V(N)$ curve on the scales appropriate to large 
scale structure (LSS), assuming $\ln V(N)$ is linear. If its second 
derivative is significant, then the maximum allowed gradient becomes less, 
vanishing completely if $\ln V(N)$ becomes too curved. The short dotted 
lines schematically illustrate the slopes consistent with inflation and the 
maximum slope permitted at the end of inflation if the general relativity 
limit is to be obtained. The illustrated $\ln V(N)$ curve satisfies all 
constraints except the general relativity limit at the end of inflation.} 
\end{figure} 

Fig.~1 schematically illustrates all the constraints. At 55 $e$-foldings 
from the end of inflation, the gradient of $\ln V(N)/V_0$ must be small 
enough to produce adequately scale-invariant density perturbations. If $\ln 
V(N)$ is linear, the gradient must be less than 0.2. If $\ln V(N)$ has 
significant curvature, this too contributes to the breaking of 
scale-invariance and the maximum gradient is reduced, vanishing entirely if 
${\rm d}^2 \ln V/{\rm d} N^2$ is outside the range $(-0.02,0.01)$. 
Expressing ${\rm d}^2 \ln V/{\rm d} N^2$ in terms of the slow-roll 
parameters
\begin{equation} 
\label{etaV} 
\frac{{\rm d}^2 \ln V } {{\rm d} N^2}  =  4 \epsilon ( \eta - 2 \epsilon),
\end{equation}
allows us to restrict  the values which $ {\rm d} \ln V/ {\rm d} N $ and 
${\rm d}^2 \ln V/{\rm d} N^2$ can simultaneously take, using the constraints 
on $\epsilon_{55}$ and $\eta_{55}$.

The bubble constraint dictates that at 55 $e$-foldings, $\ln V(N)/V_0$ must 
be at least 11.5, in order that the bubble constraint is satisfied; this 
requires a rapid decrease of $V$ as inflation proceeds. A linear 
extrapolation from the maximum allowed gradient, corresponding to an 
exponential potential, is insufficient to do the job, so the original 
extended inflation model (and any variant based on a constant $\omega$) is 
excluded. The $\ln V(N)$ curve must also descend respecting the condition 
for inflation, $\epsilon \lesssim 1$, which imposes a maximum gradient as 
illustrated. Finally, if general relativity is to be successfully attained 
after inflation without use of a potential for $\Phi$, then the gradient 
must be extremely shallow, as shown, at the end of inflation. A curve 
satisfying all these constraints yields a viable extended inflation model.

\section{Specific models}

\subsection{Extended intermediate inflation}

Intermediate inflation \cite{intinf} is an interesting model, because it is 
one of the rare inflation models which can give a `blue' ($n>1$) spectrum of 
density perturbations \cite{BL,MML}. Such spectra remain compatible with 
observational data \cite{MML,LLSSV}, provided $n$ is not too large. 
It can also give a scale-invariant spectrum \cite{BL}. Original versions of 
this model are not very satisfactory however, because there is no natural 
end to inflation. This can be rectified by implementing the model in our 
present framework; we can choose $\omega(\Phi)$ so as to give intermediate 
inflation in the Einstein frame, and have tunnelling to end inflation. 
Barrow and Maeda \cite{BM} found a class of such solutions, but at that time 
all the constraints on the scenario had not been discovered. 
Garc\'{\i}a-Bellido and Wands \cite{GBW} briefly indicated how to obtain the 
scale-invariant ($n=1$) intermediate inflation model in this way. The 
construction of viable models of this type was the original aim of our work; 
we now show that it cannot be realized.

The required coupling function $\omega(\Phi)$ is 
\begin{equation}
\omega(\Phi) + \frac{3}{2} =  \frac{8\pi}{m_{{\rm Pl}}^2}
	\frac{4}{\alpha^2} \beta^{2/\alpha} \Phi^{4/\alpha} \,,
\end{equation} 
where $\beta$ and $\alpha$ are dimensionless constants. From 
Eqs.~(\ref{phiPhi}) and (\ref{VPhi}), the Einstein frame potential has 
form $V(\phi) = \beta V_0 \phi^{-\alpha}$, which is the potential 
whose slow-roll solutions give intermediate inflation \cite{intinf,BL}. For 
$0 < \alpha < 2$ it produces a blue spectrum \cite{BL}.

At large $V(\phi)$, the intermediate inflation potential is too steep to 
support inflation. Assuming that inflation commences when the 
potential becomes flat enough that $\epsilon = 1$, Eq.~(\ref{N}) leads to an 
expression for $\phi$ at the end of inflation in terms of $\alpha$ and the 
total number of $e$-foldings $N_{{\rm tot}}$, which is
\begin{equation}
\phi_{{\rm end}}^2 = \frac{\alpha m_{{\rm Pl}}^2}{16 \pi} 
	\left(4 N_{{\rm tot}} + \alpha \right) \,.
\end{equation}                  
Using this to evaluate Eq.~(\ref{neq2}) at the end of inflation produces, 
in the approximation that $N_{{\rm tot}} \gg \alpha$ and $\omega \gg 1$, a 
relationship between the final value of $\omega$, $N_{{\rm tot}}$ and 
$\alpha$, which is
\begin{equation}
\omega(\Phi=1) = \frac{8N_{{\rm tot}}}{\alpha} \,.
\end{equation}
Note that there are parameter values for which the solar system constraint 
on $\omega$ can be satisfied.

These expressions allow $V(\phi)$ to be written in a particularly simple 
form
\begin{equation}
V(N) = V_0 \left( 1-\frac{N}{N_{{\rm tot}}} \right)^{-\alpha/2} \,,
\end{equation}
so that  
\begin{equation}
\ln \frac{V(N)}{V_0} = -\frac{\alpha}{2} \ln \left( 1 - 
	\frac{N}{N_{{\rm tot}}} \right) \,.
\end{equation}
The gradient of $\ln V(N)$ increases with $N$ (that is, the curve as plotted 
in Fig.~1 curves upwards); this is a generic property of models in which 
$\omega$ is an increasing function of time. Since we have already seen that 
even a straight line (corresponding to power-law inflation) is excluded, 
this immediately means that if the amplitude of the potential is great 
enough to satisfy the bubble constraint, then the gradient evaluated over 
the range $50 < N < 60$ violates the constraints imposed by large scale 
structure. Indeed, the bubble constraint alone is sufficient to rule out 
intermediate inflation for the range of $\alpha$ values giving $n>1$.

The failure of this model immediately indicates how hard it will be to 
implement extended inflation models giving blue spectra, because from 
Eqs.~(\ref{n}), (\ref{epsV}) and (\ref{etaV}) the condition for $n>1$ is 
simply
\begin{equation}
\frac{{\rm d}^2 \ln V}{{\rm d}N^2} > \left( \frac{{\rm d} \ln V}{{\rm d}N}
	\right)^2 \,,
\end{equation}
implying that $\ln V(N)$ must be curving up at 55 $e$-foldings. This could 
only give a model satisfying all the constraints if at lower values of $N$ 
it is curving down. This can only be achieved by using a potential (or 
equivalently $\omega(\Phi)$) which has many features during the late stages 
of inflation.

\subsection{General forms for $\ln V(N)$}

To end our discussion, we show that it is possible to construct models 
satisfying all the constraints, with the exception of the present value of 
$\omega$ (which could also be achieved by a suitably modified $\ln V(N)$ 
curve) which we shall assume is salvaged by a potential for $\Phi$. We 
investigate two general forms for $\ln V(N)$, identifying the range of 
parameter values for which the bubble constraint and the large scale 
structure constraints on the curvature of $\ln V(N)$ are simultaneously 
satisfied. Both these models are quite contrived, illustrating how difficult 
it is to remain within the presently existing constraints.

Given a particular form for $\ln V(N)$, it is possible to find $V(\phi)$ 
using the following relation, derived from Eq.~(\ref{N}), to obtain 
$\phi(N)$
\begin{equation}
\phi - \phi_{{\rm end}} = \frac{m_{\rm {{Pl}}}}{\sqrt{8\pi}} \int_0^N
	\left( \frac{{\rm d} \ln V}{{\rm d}N} \right)^{1/2} {\rm d} N \,,
\end{equation}
where, without loss of generality, we have assumed $\dot{\phi}$ 
positive. 

\begin{description}
\item[Model 1:] $\ln V(N)/V_0 = aN - bN^2$

This model satisfies all the constraints for $0.23 < a < 0.77$, with $b$ 
allowed to take a narrow range of small positive values, typically around 
$10^{-3}$, for a given $a$. The small negative quadratic term reduces 
the curvature of $\ln V(N)/V_0$ for large $N$ without significantly reducing 
its amplitude.
 
In terms of $\phi$, this potential has form
\begin{equation}
\ln \frac{V(\phi)}{V_0} = \frac{a^2}{4b} \left[1 - \left( 1 -
	\frac{\sqrt{72\pi} \, b}{a^{3/2}} \, \frac{\phi_{{\rm end}}
	-\phi}{m_{{\rm Pl}}} \right)^{4/3} \right] \,.
\end{equation} 

\item[Model 2:] $\ln V(N)/V_0 = A\sin \left( \pi N/2 N_{{\rm max}} \right)$

For this rather unnatural model, values of $A$ between 11.8 and 22.5 are 
allowed; however the values of $N_{{\rm max}}$ permitted for any given $A$ 
are strongly limited by the large scale structure constraints on the 
gradient and curvature of $\ln V(N)/V_0$. For $11.8 < A < 16.5$ the region 
of allowed values of $N_{{\rm max}}$ forms a single band, with a lower 
limit of 63. However for $16.5 < A < 20 $ this splits into two narrow bands 
and finally for $20 < A < 22.5$ only a small range of values around  
$N_{{\rm max}} \approx 145$ are allowed.

\end{description}

\section{Conclusions}

We have carried out a model-independent analysis of the constraints on 
extended inflation scenarios containing a trapped scalar field within an 
arbitrary scalar--tensor theory. We have shown that all of the constraints 
can be interpreted graphically, by considering the behaviour of the 
(logarithm of the) Einstein frame potential as a function of the number of 
$e$-foldings from the end of inflation. The principal competition arises 
from the need to keep the density perturbation spectrum adequately 
scale-invariant while suppressing the production of bubbles which finish 
with astrophysically large sizes.

Especially bearing in mind that our implementation of the constraints is 
quite conservative, we have been able to show how difficult it is to obtain 
a successful extended inflation scenario. Indeed, all models of this 
type which presently exist in the literature, in which bubble nucleation 
ends inflation, cannot evade the combination of constraints. Things seem 
particularly tough if one desires a `blue' spectrum of perturbations, which 
is a situation in which one would have hoped extended inflation might have 
fared well since such models are hard to implement in the chaotic inflation 
framework. Our graphical approach allows one to see exactly what is needed 
to obtain working models; for example, any model in which $\omega$ 
monotonically increases with time will not work. We have devised models 
which are allowed, though they seem rather contrived.

To conclude, the extended inflation paradigm is an attractive one, because 
the new physics, that of extended gravitational theories, can be tested in a 
number of ways. Unfortunately, when one also adds the extra constraints 
brought on by demanding that inflation ends by a first-order transition, the 
scenario becomes so highly constrained that it is extremely hard to find any 
working models. Still, the fact that one can exclude inflationary models on 
the basis of observational data should be viewed as an encouraging 
situation, and one we shall hear much of in future years.

%%%%%%%%%%%%%%%%%%%%%%%%%%%%%%%%%%%%%%%%%%%%%%%%%%%%%%%%%%%%%%%%%%%%%%%%
\section*{Acknowledgements}

AMG is supported by PPARC and ARL by the Royal Society. We thank 
Juan Garc\'{\i}a-Bellido and David Wands for their comments, and acknowledge 
use 
of the Starlink computer system at the University of Sussex. 
%%%%%%%%%%%%%%%%%%%%%%%%%%%%%%%%%%%%%%%%%%%%%%%%%%%%%%%%%%%%%%%%%%%%%%%%
 
\appendix
\section*{Relations between the Jordan and Einstein frames}

Some aspects of the relation between the frames have been considered by 
Lidsey \cite{Lids}. We shall use the slow-roll parameters $\epsilon$ and 
$\eta$ to generalize those arguments, producing exact relationships between 
Einstein and Jordan frame quantities as well as investigating the slow-roll 
limit.

\subsection{Condition for Inflation}

Inflation corresponds to an accelerating scale factor. However, it is 
possible for the scale factor to be accelerating in one frame and not the 
other. The inflationary condition can be rewritten as
\begin{equation}
\frac{{\rm d} H}{{\rm d} t} + H^2 > 0 \,, 
\end{equation}
(and similarly in the Einstein frame) which is more convenient for 
comparison between the Einstein and Jordan frames.

There is an exact relation given by
\begin{equation}
\frac{1}{a} \frac{{\rm d}^2 a}{{\rm d}t^2} = \Omega^{2} 
	\left(\frac{{\rm d} \tilde{H}}{{\rm d} \tilde{t}} +
	\tilde{H}^2 \right) - \Omega \frac{{\rm d} \Omega}{{\rm d}
	\tilde{t}} \tilde{H} + \left( \frac{{\rm d} \Omega}{{\rm d}
	\tilde{t}}\right)^2 - \Omega \frac{{\rm d}^2 \Omega}{{\rm d}
	\tilde{t}^2} \,,
\end{equation}
and using Eq.~(\ref{omeps}) we obtain
\begin{eqnarray}
\frac{1}{a} \frac{{\rm d}^{2}a}{ {\rm d}t^{2}} &  =  & \Phi \left[  
	\left( \frac{ {\rm d} \tilde{H}}{{\rm d} \tilde{t}} + 
	\tilde{H}^2 \right) -\frac{ \sqrt{\pi \epsilon}}{m_{{\rm Pl}}}
	\left( \frac{{\rm d} \phi}{{\rm d} \tilde{t}} \tilde {H} 
	- \frac{{\rm d}^{2} \phi}{{\rm d} \tilde{t}^{2}} \right) 
	\right] \nonumber \\
 & &  - \Phi \left[ \frac{ 8 \pi}{m_{{\rm Pl}}^{2}} \frac{\eta - 
 	2\epsilon}{8} \left( \frac{{\rm d} \phi}{{\rm d} 
 	\tilde{t}} \right)^2 \right] \,.
\end{eqnarray}
The slow-roll approximation allows us to systematically discard terms, and 
leads, to first order in the slow-roll parameters, to the relation
\begin{equation}
\frac{1}{a} \frac{{\rm d}^{2}a}{ {\rm d}t^{2}} = \Phi \left[ \left( 
	\frac{ {\rm d} \tilde{H}}{{\rm d} \tilde{t}} + \tilde{H}^{2} 
	\right) + \frac{ 2 \pi}{3 m_{{\rm Pl}}^{2}} \, 
	\epsilon V \right] \,.
\end{equation}
The right hand side of this equation is always positive, so inflation in 
the Einstein frame guarantees inflation in the Jordan frame, as found by 
Lidsey \cite{Lids}.

\subsection{Number of $e$-foldings}

The number of $e$-foldings of inflation in the two frames, given by 
\begin{equation}
N = \int H {\rm d} t \quad ; \quad \tilde{N}  = \int \tilde{H} {\rm d} 
\tilde{t},
\end{equation}
are in general different. However, from Eqs.~(\ref{neq1}) and 
(\ref{neq2}) we can see that they are approximately equal provided
\begin{equation}
\left| \frac{{\rm d} \Omega/{\rm d} \tilde{t}}{\Omega \tilde{H}} \right|
	\ll 1 \,.
\end{equation}
With the use of Eqs.~(\ref{phiPhi}), (\ref{eps}) and (\ref{omeps}), this 
becomes simply
\begin{equation}
\epsilon \ll 2 \,.
\end{equation}
Hence in the slow-roll limit the amount of inflation as measured in the 
different frames coincides.

%%%%%%%%%%%%%%%%%%%%%%%%%%%%%%%%%%%%%%%%%%%%%%%%%%%%%%%%%%%%%%%%%%%%%%%%

\end{document}